\documentclass[12pt,preprint]{aastex}
\shorttitle{{\it XMM} Observations of BAT AGN}
\shortauthors{Trippe et al.}
\usepackage{natbib}
\usepackage{subfig} 
\bibpunct[ ]{(}{)}{;}{a}{}{,}
\begin{document}
\title{{\it XMM} Follow-Up Observations of Three {\it Swift} BAT-Selected Active Galactic Nuclei}
\author{M. L. Trippe\footnotemark[1], C. S. Reynolds\footnotemark[1], M. Koss\footnotemark[1], R. F. Mushotzky\footnotemark[1], and L. M. Winter$^{2, 3}$ }
\footnotetext[1]{Department of Astronomy, University of Maryland, College Park, MD, 20742}
\footnotetext[2]{Center for Astrophysics and Space Astronomy, University of Colorado, Boulder, CO 80309}
\footnotetext[3]{Hubble Fellow}
\begin{abstract}
\indent We present {\it XMM-Newton} observations of three AGN taken as part of a hunt to find very heavily obscured Compton-thick AGN. For obscuring columns greater than 10$^{25}$ cm$^{-2}$, AGN are only visible at energies below 10 keV via reflected/scattered radiation, characterized by a flat power-law. We therefore selected three objects (ESO 417-G006, IRAS 05218-1212, and MCG -01-05-047) from the {\it Swift} BAT hard X-ray survey catalog with {\it Swift} X-ray Telescope (XRT) 0.5-10 keV spectra with flat power-law indices as candidate Compton-thick sources for follow-up observations with the more sensitive instruments on {\it XMM-Newton}. The {\it XMM} spectra, however, rule out reflection-dominated models based on the weakness of the observed Fe K$\alpha$ lines. Instead, the spectra are well-fit by a model of a power-law continuum obscured by a Compton-thin absorber, plus a soft excess. This result is consistent with previous follow-up observations of two other flat-spectrum BAT-detected AGN. Thus, out of the six AGN in the 22-month BAT catalog with apparently flat {\it Swift} XRT spectra, all five that have had follow-up observations are not likely Compton-thick. \\
\indent We also present new optical spectra of two of these objects, IRAS 05218-1212 and MCG -01-05-047. Interestingly, though both these AGN have similar X-ray spectra, their optical spectra are completely different, adding evidence against the simplest form of the geometric unified model of AGN. IRAS 05218-1212 appears in the optical as a Seyfert 1, despite the $\sim8.5\times10^{22}$ cm$^{-2}$ line-of-sight absorbing column indicated by its X-ray spectrum. MCG -01-05-047's optical spectrum shows no sign of AGN activity; it appears as a normal galaxy. \\
\end{abstract}
\section{Introduction}
\indent The most difficult to detect nearby active galactic nuclei (AGN) are those that are the most highly obscured; little or none of the emission in the optical, UV, or soft X-ray that is typically used to identify an active nucleus may be observable in Compton-thick AGN, those AGN obscured by column densities of N$_{H}\gtrsim10^{24}$ cm$^{-2}$ \citep[for reviews of Compton-thick AGN, see][]{com04,del08}. The number density of these elusive AGN is an important parameter in understanding the history of black hole growth \citep{mar04} and feedback/galaxy formation processes \citep{dim05} as well as the X-ray background \citep{gil07,tre09,dra09}, but as-yet it has not been well-determined observationally. \\
\indent Hard X-ray surveys above $\sim10$ keV offer a unique opportunity to determine the local number density of highly obscured AGN, as these wavelengths are less affected by the strong biases from obscuration and dilution by starlight that hamper the detection of these objects at shorter wavelengths. The {\it Swift} Burst Alert Telescope (BAT) has performed a whole-sky survey in the 14-195 keV waveband, and the 22-month BAT catalog has identified $\sim$270 AGN \citep{tue10}. {\it Swift} also has the capability to perform pointed X-ray observations in the 0.5-10 keV band with its X-ray Telescope \citep[XRT;][]{bur05}, and has obtained follow-up observations of $\sim180$ of the BAT-selected AGN. \\
\indent The appearance of the 0.5-10 keV spectra of Compton-thick AGN depends upon the degree to which they are obscured. If the line-of-sight obscuration is only mildly optically thick to Compton scattering (N$_{H}$ less than a few times $10^{24}$ cm$^{-2}$), the highly absorbed continuum can still be observed at energies less than 10 keV. However, for the most highly obscured AGN (N$_{H}$ greater than $10^{25}$ cm$^{-2}$), the direct continuum will be completely suppressed below 10 keV and the AGN will instead be visible in this regime only by X-rays scattered/reflected back into our line of sight. Reflection of a $\Gamma=1.9$ power-law off a slab of neutral gas (ionization parameter $\xi<10$ erg cm s$^{-1}$) produces a spectrum characterized by $\Gamma<0.5$ in the 0.5-10 keV regime. AGN continuum reflection from the inner wall of an optically thick parsec-scale torus (like those proposed in the unification model) will therefore have a flat power-law continuum, along with a strong neutral Fe K$\alpha$ fluorescent emission line at 6.4 keV \citep[for further information and in-depth simulations of the X-ray spectra of AGN surrounded by Compton-thick tori, see][]{mur09}. Because of their different paths to our line of sight, the reflection spectrum and the primary continuum pass through different absorbers, with the result that the column density measured by spectral fitting of the 0.5-10 keV band may not be indicative of heavy obscuration hiding the primary continuum. \\
\indent In order to try to find heavily Compton-thick AGN in the BAT survey that could have escaped identification as highly obscured objects due to their apparently low columns, we searched the internal {\it Swift} team database of XRT spectra of the 22-month survey catalog for sources that seemed to be dominated by reflected light, as evidenced by very flat power-law continua. Specifically, we selected sources that 1.) had a best-fitting photon index less than $\Gamma=0.5$, 2.) were of sufficient quality to impose a 1$\sigma$ upper bound on the photon index of less than $\Gamma=1$, 3.) had a known optical counterpart and well-determined redshift, and 4.) were not flagged as blazars on the basis of their optical or radio properties. We found six AGN met these requirements; two already had follow-up observations to the XRT data made by the more sensitive instruments on board {\it XMM-Newton} \citep[Mrk 417 and NGC 612;][]{win08}. In this paper we present follow-up X-ray spectroscopy with {\it XMM} of three of the remaining objects, ESO 417-G006 (XRT $\Gamma=-0.59^{+1.12}_{-1.17}$), IRAS 05218-1212 (XRT $\Gamma=-0.43^{+0.32}_{-0.34}$), and MCG -01-05-047 (XRT $\Gamma<-1.5$). The final remaining object, 2MASX J03565655-4041453, was also observed with {\it XMM} as part of this program but the observation was lost due to background flares. We also present new optical spectra of two of these objects, IRAS 05218-1212, and MCG -01-05-047. \\
\section{X-ray Observations}
\subsection{Data Analysis}
\ The {\it XMM-Newton} satellite carries three X-ray CCD cameras: two metal oxide semiconductor CCDs known as the MOS cameras, and one pn-type CCD known as the pn camera. These CCDs allow observations in the range $\sim$0.5-10 keV, with an energy resolution full width at half maximum (FWHM) of $\sim$130 eV at 6 keV. Through the {\it XMM} Guest Observer program, we obtained observations of ESO 417-G006, IRAS 05218-1212, and MCG -01-05-047. The dates of the observations and the exposure times are given in Table~\ref{logtbl}, the observation log. \\
\indent The data were reprocessed with the standard EPCHAIN and EMCHAIN processing scripts included in version 9.0 of the XMM-SAS (Science Analysis System) software, using the most recent calibration files in the CALDB. The data were then filtered to exclude times of high background noise, and source spectra were extracted from circular regions 32$''$ in diameter. Background spectra were also extracted in 32$''$ circles from the same chip as the source spectra, but from areas free of any background objects. Response matrices and ancillary response matrices were generated using the SAS tasks {\it rmfgen} and {\it arfgen}. \\
\indent For ESO 417-G006 and MCG -01-05-047 we used the pn spectrum for spectral fitting, but the observation of IRAS 05218-1212 fell into a gap between the pn chips and we therefore use the combined MOS1 and MOS2 spectrum instead of the pn spectrum for this source. All fits were performed using XSPEC version 12.5.1. \\
\indent In addition to the {\it XMM} data, we downloaded the archival 22-month 8-channel {\it Swift} BAT spectra of these objects from the HEASARC to include in our analysis. Details of the processing of these spectra are given in \citet{tue10}. Because reflection of the direct power-law continuum off optically thick material is expected to produce a broad ``Compton reflection'' hump in the $15-100$ keV continuum, these data are a useful complement to the lower-energy {\it XMM} data when trying to determine if the spectrum is best fit by a reflection model \citep{sev11}. \\
\subsection{Spectral Models}
\begin{figure}%
\centering
\subfloat[ESO 417-G006]{\label{esolc}\includegraphics[angle=90,scale=.32]{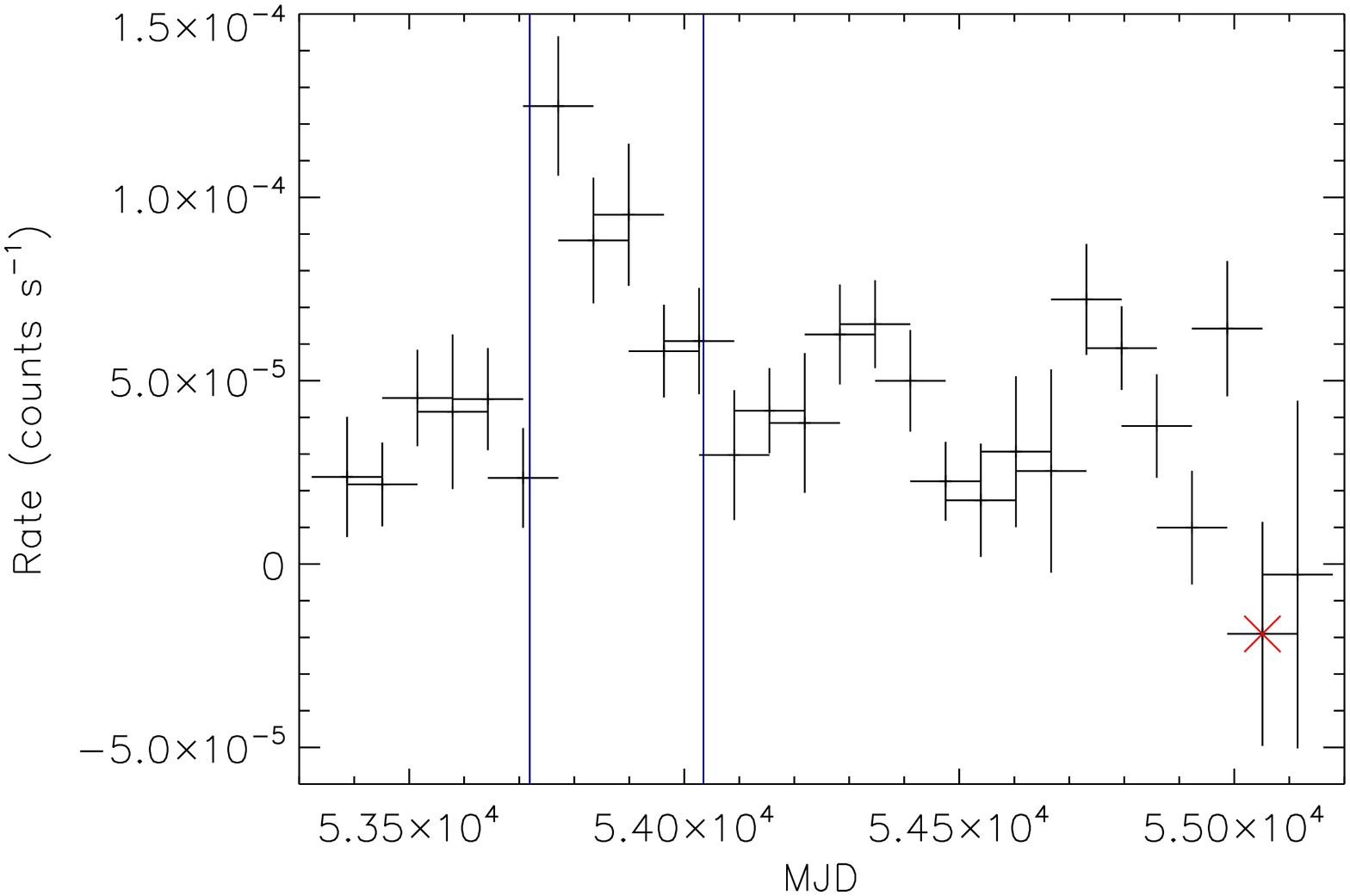}}\qquad
\subfloat[IRAS 05218-1212]{\label{iraslc}\includegraphics[angle=90,scale=.32]{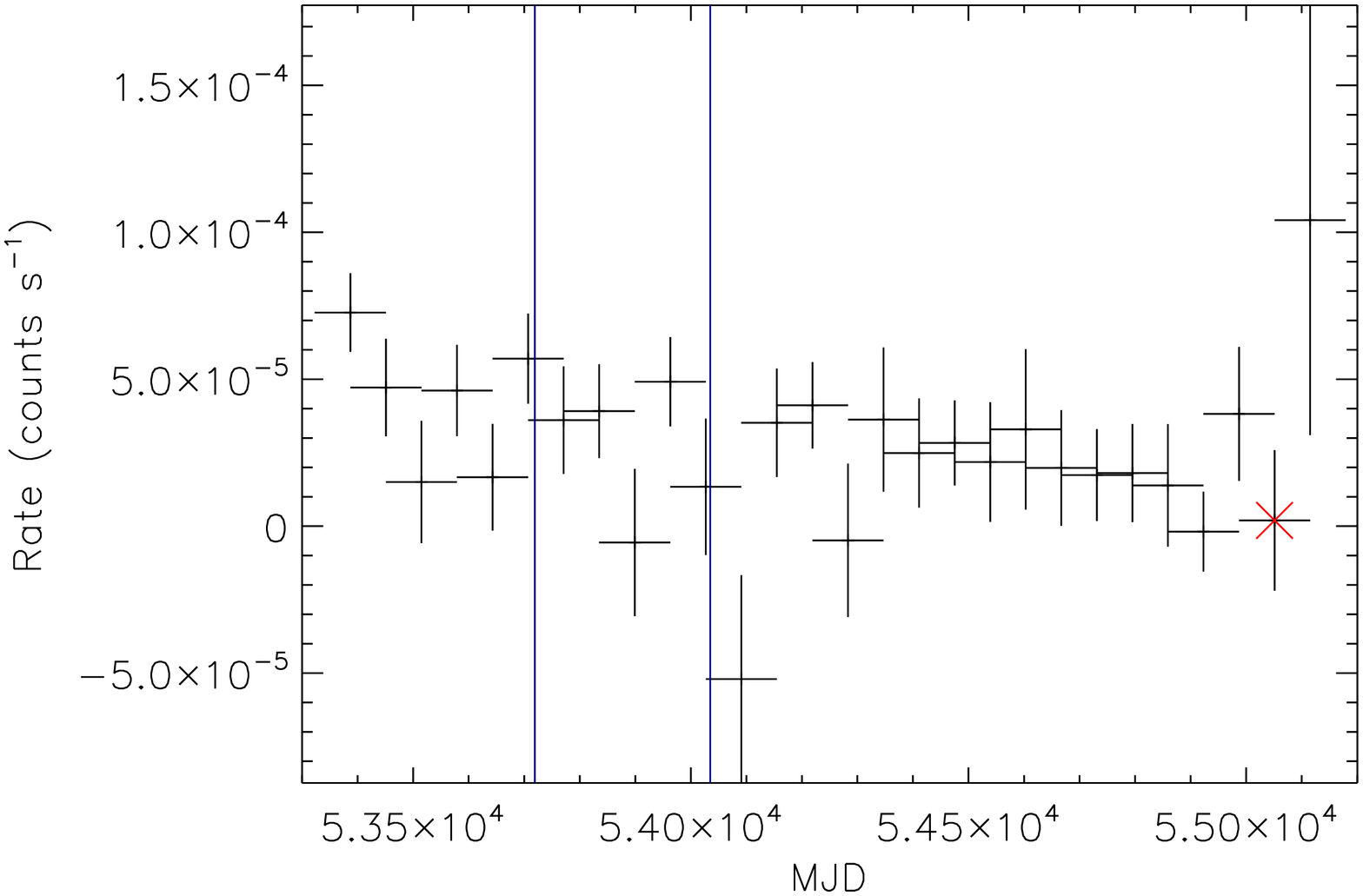}}\\
\subfloat[MCG-01-05-047]{\label{mcglc}\includegraphics[angle=90,scale=.32]{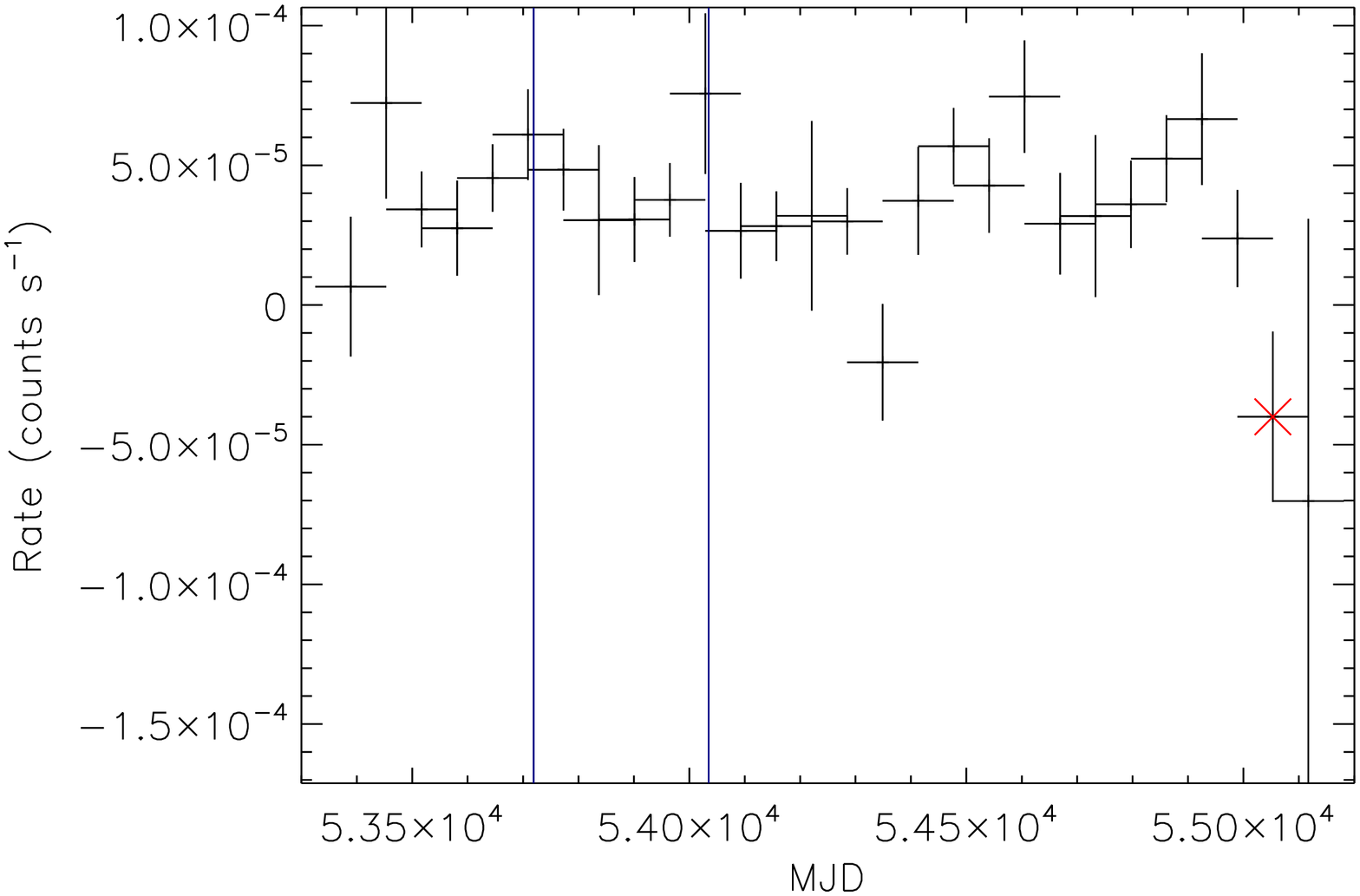}}%
\caption{58-Month BAT 14-195 keV lightcurves for ESO 417-G006, IRAS 05218-1212, and MCG-01-05-047, binned to 64 day intervals. The two vertical lines denote the start and end of the 22-month BAT survey, and the red X marks the time of observation with {\it XMM}. }
\label{lcfig}
\end{figure}

\begin{figure}
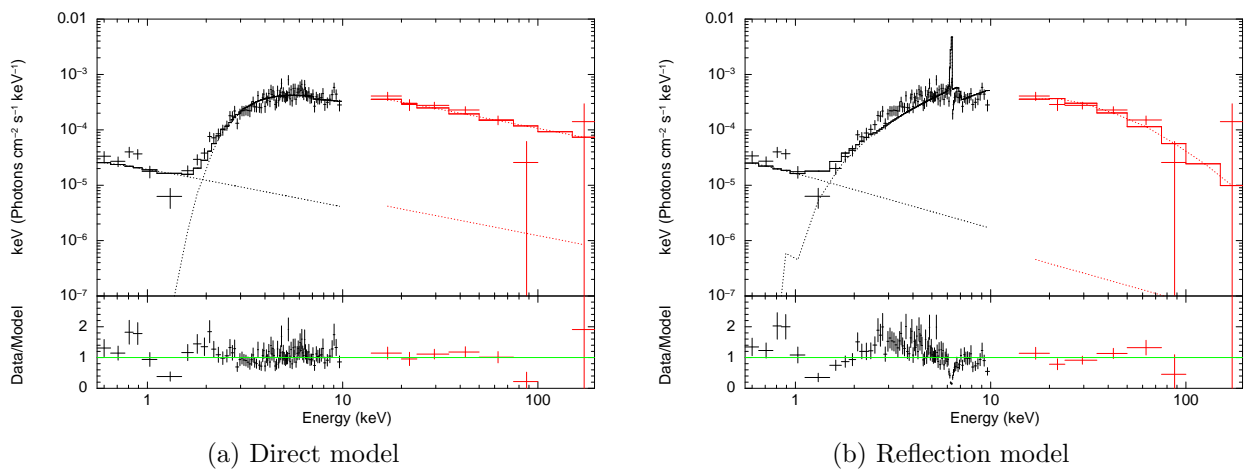
%
\centering
\subfloat[Direct model]{\label{esomod1}\includegraphics[angle=270,scale=.31]{eso_dir_rat.ps}}\qquad
\subfloat[Reflection model]{\label{esomod2}\includegraphics[angle=270,scale=.31]{eso_refl_rat.ps}}%
\caption{ESO 417-G006 {\it XMM} (black) and BAT (red) data, fit with (a) a ``direct'' model comprised of a power-law continuum modified by absorption plus a power-law soft excess, and (b) a pure reflection model, using {\it reflionx}.  Note that the y-axes of the bottom panels are scaled to best show the residuals in the {\it XMM} pn data; a couple of the BAT points with very large error bars are off this scale. }
\label{esofig}
\end{figure}

\begin{figure}%
\centering
\subfloat[Direct model]{\label{irasmod1}\includegraphics[angle=270,scale=.31]{iras_dir_rat.ps}}\qquad
\subfloat[Reflection model]{\label{irasmod2}\includegraphics[angle=270,scale=.31]{iras_refl_rat.ps}}\\
\subfloat[Optical spectrum]{\label{irasopt}\includegraphics[angle=90,scale=.35]{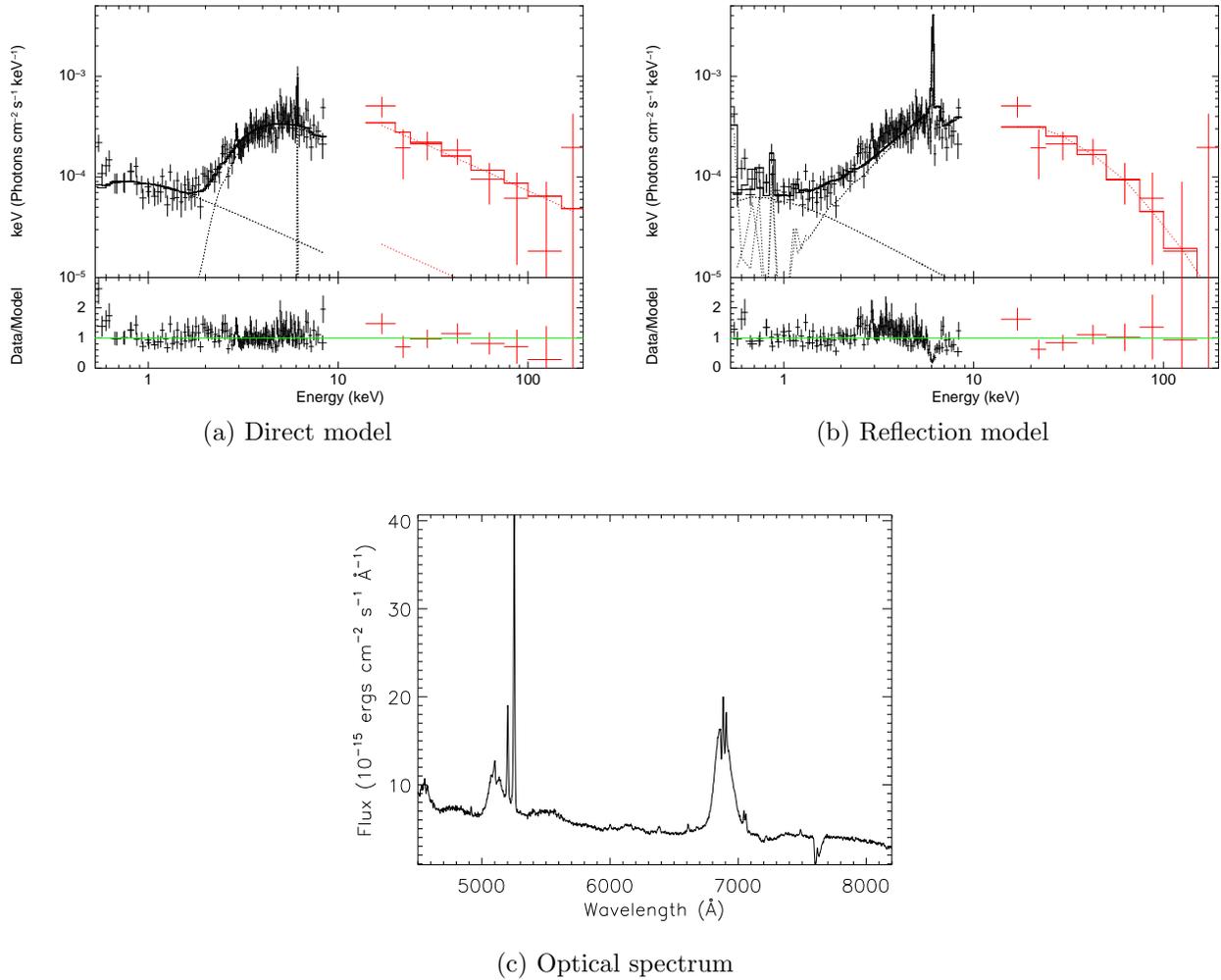}}%
\caption{IRAS 05218-1212 {\it XMM} (black) and BAT (red) data, fit with a ``direct'' model (a) and a pure reflection model (b). Plot (c) shows IRAS 05218-1212's optical spectrum, typical of a type 1 Seyfert.}
\label{irasfig}
\end{figure}

\begin{figure}%
\centering
\subfloat[Direct model]{\label{mcgmod1}\includegraphics[angle=270,scale=.31]{mcg_dir_rat.ps}}\qquad
\subfloat[Reflection model]{\label{mcgmod2}\includegraphics[angle=270,scale=.31]{mcg_refl_rat.ps}}\\
\subfloat[Optical spectrum]{\label{mcgopt}\includegraphics[angle=90,scale=.35]{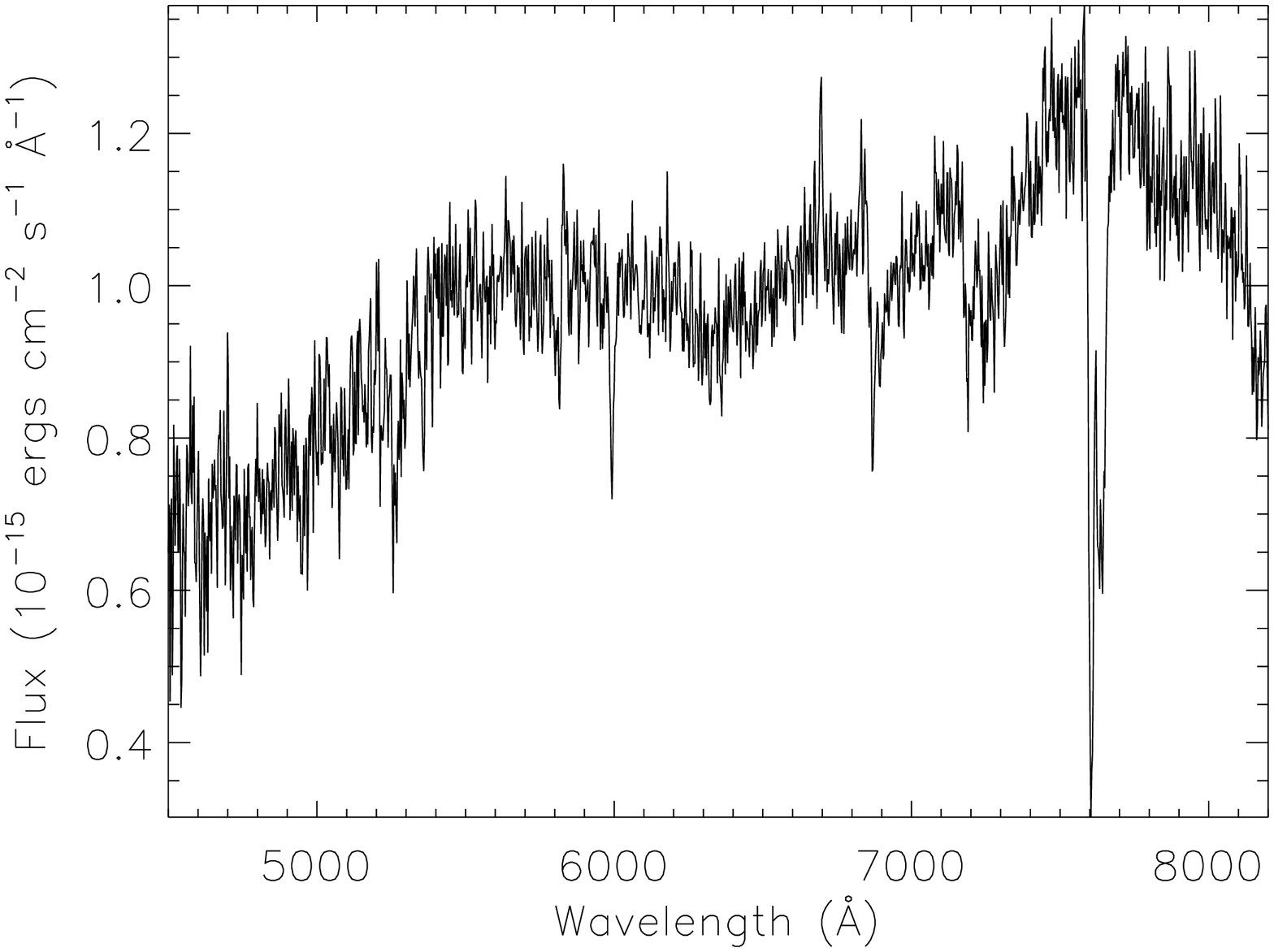}}%
\caption{MCG-01-05-047 {\it XMM} (black) and BAT (red) data, fit with a ``direct'' model (a) and a pure reflection model (b). Note that the y-axes of the bottom panels are scaled to best show the residuals in the {\it XMM} pn data; a couple of the BAT points with very large error bars are off this scale. Plot (c) shows MCG-01-05-047's optical spectrum, which lacks the usual emission lines used to identify AGN in the optical.}
\label{mcgfig}
\end{figure}

\indent To determine if the X-rays from these AGN are direct or reflected, we fit two models to the data for each object and compared them to determine the best fit.  In order to fully characterize the data, the {\it XMM} and BAT data were fit simultaneously. To investigate the possibility of variability in the BAT band between the BAT spectra (average spectra over the first 22 months of the survey) and the later {\it XMM} observations, we downloaded the publicly-available 58-month BAT lightcurves (Baumgartner et al. 2011, submitted to ApJS) of each object, binned to 64 day intervals (see Figure~\ref{lcfig}). The lightcurve of ESO 417-G006 shows some evidence for a factor of $\sim$2 increase in count rate during the 22-month survey, while the lightcurves of IRAS 05218-1212 and MCG -01-05-047 do not present obvious evidence for variability.  Because the cross-normalization factor between the instruments is not known, and to account for the variability in the BAT flux of ESO 417-G006, in all of our joint {\it XMM} and BAT fits we allow the BAT flux to vary by a constant factor. Of course, this accounts for variability in intensity only and not spectral shape, and therefore the broad-band form of the continuum of ESO 417-G006 is still somewhat uncertain. All error bars are quoted at the $90\%$ confidence level. \\
\indent The first model we fit to the data, the ``direct'' model, consisted of a primary power-law continuum modified by absorption, plus a Gaussian Fe K$\alpha$ line, {\it constant*tbabs(powerlaw +tbabs*powerlaw +zgaussian)} in XSPEC. We used a secondary powerlaw with index tied to that of the primary, to model the ``soft excess'' seen in all of the spectra in the 0.5-2 keV band. Thus, we represent the soft excess as continuum light that has scattered around the absorber, though our data does not allow us to unambiguously determine that this is the case. It could alternatively be emission from a starburst component (though we note that the optical spectra do not show evidence for strong starbusts) or emission from extended circumnuclear gas photoionized by the AGN \citep{gua07}. The absorption model, {\it tbabs}, was used with solar abundances from \citet{wil00} and photoionization cross-sections from \citet{ver96}. The column density of the first {\it tbabs} component in each model was frozen to the Galactic absorption value measured by \citet{dic90}. The second {\it tbabs} column, representing absorption intrinsic to the source, was allowed to vary. Table~\ref{pcfparamtable} gives the parameters of the best-fit model for each AGN, and Figures~\ref{esomod1}, \ref{irasmod1}, and \ref{mcgmod1} show the fits. \\
\indent This model provided good fits to the data; reduced $\chi^{2}$ values of 1.45, 1.00, and 1.34 were found for ESO 417-G006, IRAS 05218-1212, and MCG -01-05-047, respectively. In IRAS 05218-1212 and MCG -01-05-047, unresolved Fe K$\alpha$ emission was detected at the 99\% significance level, while in ESO 417-G006 no line was detected at the 90\% level. MCG -01-05-047 was the most highly obscured of the three ($N_{H}=26.3_{-1.0}^{+1.0}\times10^{22}$ cm$^{-2}$), ESO 417-G006 next most highly ($N_{H}=13.5_{-1.4}^{+1.6}\times10^{22}$ cm$^{-2}$), and IRAS 05218-1212 the least ($N_{H}=8.5_{-0.8}^{+0.8}\times10^{22}$ cm$^{-2}$). ESO 417-G006 also showed the lowest photon index of the three, $\Gamma=1.70$, while for IRAS 05218-1212 $\Gamma=1.85$, and the best-fit for MCG -01-05-047 was found with $\Gamma$ frozen to 2, because of its large absorption. \\
\indent The scattering fraction (the ratio of the 0.5-2 keV soft power-law luminosity to the total absorption-corrected 0.5-2 keV model luminosity; see \citet{nog10}) of IRAS 05218-1212 was found to be about 6\%. ESO 417-G006 and MCG -01-05-047 had lower scattering fractions of only about 1\%. While optically selected Seyfert 2s typically show scattering fractions of 3-10\% \citep{tur97,cap06}, \citet{win09a} found that many (24\%) of the 9-month sample of BAT-selected AGN show scattering fractions of $<3$\%, presumably due to the BAT's unbiased sampling of Compton-thin sources, and thus the low scattering fractions of ESO 417-G006 and MCG -01-05-047 are not surprising.\\ 
\indent We then fit the data with a reflection model, {\it constant*tbabs*(powerlaw+tbabs*reflionx)}, to investigate the possibility that the spectra are reflection dominated. The powerlaw represents continuum light that scatters around or leaks through the presumed very Compton-thick absorption blocking the primary continuum, while the {\it reflionx} model represents the spectrum reflected by an optically thick slab, and includes a self-consistent value for Fe K$\alpha$ line emission \citep[see][]{ros05}. To simulate reflection off neutral material, we froze the ionization parameter $\xi$ to its minimum value (1.0) in all models. We also initially froze the value of the iron abundance to the solar value. The first {\it tbabs} column was again frozen to the Galactic value, while the second {\it tbabs} was allowed to vary to account for additional absorption of the reflection spectrum by material intrinsic to the AGN. \\
\indent Overall, the {\it reflionx} models did not produce good fits, largely because the Fe K lines predicted by this model are far too strong compared to the lines in our data. When the Fe K line region was excluded and the model fit to the continuum only, the reduced $\chi^{2}$ values were only slightly larger than for the direct models. However, once the Fe K$\alpha$ region was included reduced $\chi^{2}$ increased dramatically (see Table~\ref{reflparamtable} and Figures~\ref{esomod2}, \ref{irasmod2}, and \ref{mcgmod2}). Table~\ref{ewcomp} compares the values of the equivalent width (EW) from the best-fit {\it reflionx} model with solar abundances to the measured (or upper limit to) Fe K$\alpha$ EW determined from the Gaussian Fe line of the direct model. Allowing the iron abundance to vary from solar improves the value of $\chi^{2}$, but still does not lead to a good fit to the data. While reducing the iron abundance reduces the worst discrepancy between original model and the data (the over-strong Fe K$\alpha$ line), the underestimation of the observed continua in the 2-5 keV range becomes worse, and therefore reduced $\chi^{2}$ remains $\gtrsim$2 even in the extreme and probably unphysical case where the iron abundance is only 0.1 solar. \\
\indent Comparison of these two initial models shows that the spectra are not likely to be reflection-dominated, though both models over-simplify the true situation. Assuming that the observed column densities correspond to absorption by a simple spherical shell of neutral material surrounding the continuum predicts weaker Fe K$\alpha$ emission \citep[$\sim$40 eV for IRAS 05218-1212 and $\sim$80 eV for MCG -01-05-047;][]{lea93} than is observed ($\sim$180 eV for IRAS 05218-1212 and $\sim$166 eV for MCG -01-05-047). In reality the spectra likely have both direct and reflected components, each affected by several layers of absorption with different ionization parameters. In order to get a more physical picture, we next fit the spectra with the MYTorus model of \citet{mur09}, which provides a self-consistent calculation of the reflected and transmitted continua and line emission of a power-law continuum source surrounded by a neutral torus of obscuring material with a $60^{\circ}$ half-opening angle. These fits reinforced the general conclusions from the simpler models; the best-fit models show a strong transmitted continuum component absorbed by Compton-thin obscuration, with columns similar to those from the direct models. By setting the inclination angle of the torus to different values and re-fitting, the MYTorus models constrain the strength of the reflection component to be only $\sim3-10$\% of the primary continuum in the K$\alpha$ region in each of the three sources.
\section{Optical Observations}
\indent We were also able to obtain optical spectra for two of the objects, IRAS 05218-1212 and MCG-01-05-047, on 22 April 2009 using the Goldcam spectrograph on the 2.1-m telescope at Kitt Peak National Observatory in Arizona. Spectra were obtained with a grating with a resolution of $\sim400$ km s$^{-1}$(2.47 \AA\ $/$pixel dispersion) in the wavelength range of approximately 4300-8000 \AA\, and and a Schott GG 420 filter. For each observation, we used a slit with a 2'' width at a 90$^{\circ}$ P.A. centered on the galaxy's nucleus. To eliminate cosmic ray hits, we took three exposures each time the galaxy was observed. The stars LTT 4364 and Feige 110 were also observed with these setting for the purpose of flux calibration. The spectra were then reduced and flux calibrated using standard IRAF reduction packages for long-slit spectroscopy, and corrected for Galactic extinction using the IDL procedure CCM\_UNRED from the Goddard IDL Astronomy User's Library with extinction values from NED. The spectra are shown in Figure~\ref{irasopt} (IRAS 05218-1212) and Figure~\ref{mcgopt} (MCG-01-05-047). \\
\indent Due to its low declination, we were not able to obtain our own spectrum of ESO 417-G006 at Kitt Peak. However, an optical spectrum of this object is presented in \citet{fra03}. \\
\section{Notes on the Spectra}
\subsection{ESO 417-G006}
\indent ESO 417-G006 is classified as a Seyfert 2, though depending upon the classification criteria it could alternatively be considered a LINER, due to the very strong [O~II] $\lambda$3727 line in its optical spectrum \citep[see][]{fra03}. Its X-ray spectrum shows obscuration by a column of $\sim13.5\times10^{22}$ cm$^{-2}$, typical of Seyfert 2 galaxies \citep{ris99}. The spectrum does not show significant Fe K$\alpha$ emission, and we measure the 90\% upper limit on the EW of the line to be only 0.112 keV. \\
\subsection{IRAS 05218-1212}
\indent IRAS 05218-1212 is an unusual object: though it is obscured by a column of $\sim8.5\times10^{22}$ cm$^{-2}$, similar to that of many Seyfert 2s, its optical spectrum is dominated by the strong broad lines and blue continuum characteristic of Seyfert 1s. The addition of an unresolved Gaussian line to the direct model continuum resulted in a $\Delta \chi^{2}$ significant at the 99\% level. However, because this line has an EW of only 0.180 keV, reflection is not likely to be the dominant spectral component. \\
\indent There are a couple of explanations for the apparent X-ray/optical mismatch in IRAS 05218-1212. Because the optical and X-ray spectra are not simultaneous, we cannot rule out the possibility that we have caught the AGN in two different absorption states. However, it seems unlikely that variability is the cause of the discrepancy in IRAS 05218-1212, as it has been consistently classified as an optical type 1 over the history of its observation \citep{ost85,mor89,deg92}. An EXOSAT study conducted within two years of the optical spectroscopy of \citeauthor{ost85} found it to be a factor of 10 under-luminous in the 2-10 keV X-ray band compared with similar IRAS-selected Seyferts \citep{war88}. Furthermore, our own X-ray and optical data were taken only 8 months apart, even closer in time than these previous data. If the absorption is not variable, it may be that some of the obscuring material is relatively dust free, either residing within the dust sublimation radius or originating from that region, such that it absorbs the soft X-rays but does not add to the extinction in the optical. Alternatively, it may be that the obscuring material is clumpy \citep[see, e.g.,][for a review]{eli08}, and the X-ray absorber causing a high line-of-sight column does not significantly obscure the more extended optical line-emitting gas. \\
\indent Interestingly, several large surveys have hinted that this type of X-ray/optical mismatch may not be uncommon \citep[e.g.,][]{bru03,toz06,tro09}. IRAS 05218-1212 is nearby, and therefore relatively bright compared with the higher-redshift survey AGN; it therefore warrants further study as an opportunity to gain insight into the survey results. \\
\subsection{MCG -01-05-047}
\indent MCG -01-05-047 has an X-ray spectrum typical of a Seyfert 2 galaxy (N$_{H}\sim26.3\times10^{22}$ cm$^{-2}$). Unresolved Fe K$\alpha$ emission is detected, with an EW of 0.166 keV. Interestingly, its optical spectrum is that of a normal galaxy, thus making it an example of an X-ray bright optically normal galaxy \citep{com02}. At a redshift of only 0.017, dilution of the AGN emission by the host galaxy can effectively be ruled out. However, the 2MASS image of MCG -01-05-047 shows it to be almost exactly edge-on (the ratio of minor to major axis length $b/a=0.11$, from NED). \citet{rig06} found obscuration by dust in the host galaxy to be the main cause of the optical dullness of XBONGs, and this seems to also be the case for MCG -01-05-047. \\
\section{Conclusions}
\indent We have presented follow-up {\it XMM} observations of three AGN from the {\it Swift} BAT hard X-ray survey, selected  by their low 0.5-10 keV XRT power-law indices as candidate Compton-thick objects. The {\it XMM} spectra are well-fit with a ``direct'' model (absorbed power-law plus soft excess). Reflection-dominated models are ruled out by the weakness of the observed Fe K$\alpha$ emission line, and therefore it seems likely that none of these objects are obscured by Compton-thick material in the line of sight. In fact, fitting the spectra with the MYTorus model of \citet{mur09} indicates that only a few percent of their X-rays are from a reflection component. \\
\indent Previous follow-up observations of two other low-$\Gamma$ Compton-thick candidates \citep[Mrk 417 and NGC 612, see][]{win09b,win08}, confirm that they too are probably not Compton-thick. Thus, out of the six 22-month BAT AGN with flat XRT spectra, all five that have had follow-up observations seem not to be Compton-thick. \\
\indent While all three of our targets suffer from substantial Compton-thin obscuration, they all also show appreciable soft excess emission above the extrapolation of the hard X-ray component. This soft component apparently accounts for their appearance as flat power-law sources in the XRT data, by partially filling in the nuclear continuum's absorption trough. Our results therefore underscore the fact that the innate complexity of AGN X-ray spectra can undermine the predictive power of low signal-to-noise data; techniques to determine intrinsic absorption that rely on simplifying assumptions (hardness ratios, for example) may often be unreliable \citep[e.g.,][]{gho07}. \\
\indent Though the BAT is currently the least biased X-ray instrument for the detection of obscured AGN, it still suffers significant biases against the detection of objects in the truly Compton-thick regime. For example, using the MYTorus model to estimate the fraction of source flux lost as obscuration becomes Compton-thick shows that a source obscured by $N_{H}=10^{24}$ cm$^{-2}$ has only $\sim50$\% of its intrinsic flux is visible in the BAT band; for $N_{H}=10^{25}$ cm$^{-2}$ only $\sim5$\% of the intrinsic flux is visible \citep{bur11}. The BAT is therefore only able to detect the Compton-thick AGN with the largest apparent brightnesses, those that are very nearby or very intrinsically bright. \citet{bur11} report that the 36-month BAT survey has thus far identified nine Compton-thick AGN, but only two of these were not previously known to be Compton-thick.  The two new sources (Swift J0601.9-8636, \citealt{ued07}, and CGCG 420-015, \citealt{sev11}) are only mildly Compton-thick; the BAT has not yet discovered any previously unknown heavily Compton-thick sources. \citet{bur11} show that despite the low number of observed Compton-thick objects in the BAT survey, once the sample is corrected for its bias against the detection of heavily obscured sources, the estimated true fraction of local Compton-thick AGN is close to that predicted by cosmic X-ray background models \citep[similar results are found from the INTEGRAL survey;][]{mal09}. Although we hoped our current study would be able to unearth new examples of very heavily Compton-thick AGN, given the strong BAT bias against such objects it is not particularly surprising that we did not. \\
\indent Interestingly, despite their similar-looking X-ray spectra, all three AGN studied here appear quite different in the optical, in contradiction of the simplest form of geometric unified models of AGN. ESO 417-G006 appears as a Seyfert 2, IRAS 05218-1212 as a Seyfert 1, and MCG -01-05-047 as a normal galaxy. \\
{\it Acknowledgments} \\
\indent We thank the {\it XMM-Newton} Guest Observer program for its support under grants NNX08AX42G and NNX09AT75G, as well as the NASA Long Term Space Astrophysics grant NAG513065. \\
\indent Based on observations obtained with {\it XMM-Newton}, an ESA science mission with instruments and contributions directly funded by ESA Member States and NASA. \\
\indent This research has made use of the NASA/IPAC Extragalactic Database (NED) which is operated by the Jet Propulsion Laboratory, California Institute of Technology, under contract with the National Aeronautics and Space Administration. \\
\indent This research has made use of data obtained from the High Energy Astrophysics Science Archive Research Center (HEASARC), provided by NASA's Goddard Space Flight Center. \\

\bibliographystyle{apj} 
\bibliography{apj-jour,paper}

\clearpage
\begin{deluxetable}{lcccc}
\tabletypesize{\scriptsize}
\tablewidth{0pt}
\tablecaption{Log of Observations}
\tablehead{
\colhead{Object} & \colhead{Obs. ID} & \colhead{Date} &\colhead{Exposure} & \colhead{Ct. Rate} }
\startdata
ESO 417-G006 & 0602560201 & 11 July 2009 &6.3 ks & 0.39 (pn) \\
IRAS 05218-1212 & 0551950401 & 24 August 2008 & 28.8 ks & 0.14 (MOS1/2) \\
MCG -01-05-047 & 0602560101 &24 July 2009 & 11.2 ks &0.36 (pn)  \\
\enddata
\label{logtbl}
\end{deluxetable}

\begin{deluxetable}{rccc}
\tabletypesize{\scriptsize}
\tablewidth{0pt}
\tablecaption{Parameters for Direct Models: \protect\linebreak {\it constant*tbabs(powerlaw +tbabs*powerlaw +zgaussian) } }
\tablehead{
\colhead{Parameter} & \colhead{ESO 417-G006} & \colhead{IRAS 05218-1212} & \colhead{MCG-01-05-047} 
}
\startdata
Constant ({\it XMM})$^{a}$   & 1.0                               &  1.0                                    &  1.0                                   \\
Constant (BAT)               & 1.5$_{-0.4}^{+0.5}$               &  2.2$_{-1.0}^{+1.0}$                    &  2.1$_{-0.4}^{+0.5}$                   \\
tbabs (Galactic): N$_{H}$$^{a,b}$        & 0.019                 &  0.095                                  &  0.025                                 \\
powerlaw: $\Gamma$           & =hard powerlaw $\Gamma$           &  = hard powerlaw $\Gamma$               &  = hard powerlaw $\Gamma$              \\
powerlaw: normalization$^{d}$& 2.0$_{-0.3}^{+0.3}$$\times10^{-5}$&  1.1$_{-0.1}^{+0.1}$$\times10^{-4}$     &  3.9$_{-0.3}^{+0.3}$$\times10^{-5}$    \\
tbabs (intrinsic): N$_{H}$   & 13.5$_{-1.4}^{+1.6}$              &  8.5$_{-0.8}^{+0.8}$                    &  26.3$_{-1.0}^{+1.0}$                  \\
powerlaw: $\Gamma$           & 1.70$_{-0.15}^{+0.15}$            &  1.85$_{-0.15}^{+0.15}$                 &  2.00$^{a}$                                  \\
powerlaw: normalization$^{d}$& 1.7$_{-0.4}^{+0.6}$$\times10^{-3}$&  1.7$_{-0.4}^{+0.5}$$\times10^{-3}$     &  3.4$_{-2.0}^{+2.1}$$\times10^{-3}$    \\
zgauss: line energy$^{a,e}$  & 6.4                               &  6.4                                   &  6.4                                   \\
zgauss: $\sigma$$^{a,f}$     & 0.01                              &  0.01                                  &  0.01                                  \\
zgauss: redshift$^{a,g}$     & 0.016291                          &  0.049                                 &  0.017197                              \\
zgauss: normalization$^{h}$  & $<7.35\times10^{-5}$              & 1.26$_{-0.75}^{+0.75}$$\times10^{-5}$  &  1.01$_{-0.32}^{+0.32}$$\times10^{-5}$ \\
\hline 
$\chi^{2}/dof$          & 139.5/96           &  167.3/167            &  215.7/161           \\
Reduced $\chi^{2}$      & 1.45               &  1.00                &  1.34                \\
\enddata
\tablenotetext{a}{Indicates frozen parameter.}
\tablenotetext{b}{Frozen to Galactic value from \citet{dic90}, units 10$^{22}$ cm$^{-2}$. }
\tablenotetext{c}{Units 10$^{22}$ cm$^{-2}$.} 
\tablenotetext{d}{photons keV$^{-1}$ cm$^{-2}$ s$^{-1}$ at 1 keV}
\tablenotetext{e}{in keV}
\tablenotetext{f}{Line width in keV, here frozen to an unresolved value. }
\tablenotetext{g}{Taken from the NASA Extragalactic Database (NED, http://nedwww.ipac.caltech.edu). }
\tablenotetext{h}{Total photons cm$^{-2}$ s$^{-1}$ in the line.}
\label{pcfparamtable}
\end{deluxetable}

\begin{deluxetable}{rccc}
\tabletypesize{\scriptsize}
\tablewidth{0pt}
\tablecaption{Parameters for Reflection Models*: \protect\linebreak {\it constant*tbabs*(powerlaw+tbabs*reflionx)} }
\tablehead{
\colhead{Parameter} & \colhead{ESO 417-G006} & \colhead{IRAS 05218-1212} & \colhead{MCG-01-05-047} 
}
\startdata
Constant ({\it XMM})$^{a}$    & 1.0                 &  1.0                &  1.0                 \\
Constant (BAT)                & 0.46                &  0.46               &  0.51                \\
tbabs (Galactic): N$_{H}$$^{a,b}$& 0.019            &  0.095              &  0.025               \\
tbabs (intrinsic): N$_{H}$    & 2.9                 &  0.5                &  3.2                 \\
reflionx: Fe/solar$^{a}$      & 1.0                 &  1.0                &  1.0                 \\
reflionx: $\Gamma$$^{a}$      & 2.0                 &  2.0                &  2.0                 \\
reflionx: $\xi$$^{a,d}$       & 1.0                 &  1.0                &  1.0                 \\
reflionx: redshift$^{a,e}$    & 0.016291            & 0.049               &  0.017197            \\
reflionx: normalization$^{f}$ & 5.43$\times10^{-4}$ & 4.75$\times10^{-4}$ &  4.79$\times10^{-4}$ \\
powerlaw: $\Gamma^{a}$            & 2.0                 &  2.0                &  2.0             \\
powerlaw: normalization$^{f}$ & 1.67$\times10^{-5}$ & 7.24$\times10^{-5}$ &  3.71$\times10^{-5}$ \\
\hline 
$\chi^{2}/dof$                & 1651.2/97           &  1260.8/169          &  1280.7/162         \\
Reduced $\chi^{2}$            & 17.0                &  7.5                 &  7.91               \\
\enddata
\tablenotetext{*}{Because the reduced $\chi^{2}$ values indicate that this model is not a good fit to the data, parameter error bars were not calculated. }
\tablenotetext{a}{Indicates frozen parameter.}
\tablenotetext{b}{Frozen to Galactic value from \citet{dic90}, units 10$^{22}$ cm$^{-2}$. }
\tablenotetext{c}{Units 10$^{22}$ cm$^{-2}$.} 
\tablenotetext{d}{The ionization parameter, $\xi=4 \pi F/n$ in units erg cm s$^{-1}$, where $F$ is the total illuminating flux and $n$ is the hydrogen number density.} 
\tablenotetext{e}{Taken from the NASA Extragalactic Database (NED, http://nedwww.ipac.caltech.edu). }
\tablenotetext{f}{photons keV$^{-1}$ cm$^{-2}$ s$^{-1}$ at 1 keV}
\label{reflparamtable}
\end{deluxetable}

\begin{deluxetable}{lcc}
\tabletypesize{\scriptsize}
\tablewidth{0pt}
\tablecaption{Predicted vs. Observed Fe K$\alpha$ EW}
\tablehead{
\colhead{Object} & \colhead{EW Observed} & \colhead{EW Expected} \\
\colhead{} & \colhead{(keV)} & \colhead{(keV)}
}
\startdata
ESO 417-G006    & $<$0.112                    & 0.847 \\
IRAS 05218-1212 & 0.180$^{+0.073}_{-0.072}$   & 1.31  \\
MCG -01-05-047  & 0.166$^{+0.052}_{-0.052}$   & 1.03  \\
\enddata
\tablenotetext{*}{Comparison of the Fe K$\alpha$ EW measured in the direct model with the EW predicted by the {\it reflionx} model fit to the continuum.}
\label{ewcomp}
\end{deluxetable}

\clearpage

\clearpage

\clearpage

\clearpage

\end{document}